\documentclass{article}

\usepackage{PRIMEarxiv}

\usepackage[utf8]{inputenc} 
\usepackage[T1]{fontenc}    
\usepackage{hyperref}       
\usepackage{url}            
\usepackage{booktabs}       
\usepackage{amsfonts}       
\usepackage{nicefrac}       
\usepackage{microtype}      
\usepackage{lipsum}
\usepackage{graphicx}
\usepackage[numbers]{natbib}  
\usepackage{marvosym} 

\newcommand{\shorttitle}{VK-LSVD: A Large-Scale Industrial Dataset for Short-Video Recommendation}

\hypersetup{
pdftitle={VK-LSVD: A Large-Scale Industrial Dataset for Short-Video Recommendation},
pdfsubject={
        cs.IR,          
        cs.LG,          
        cs.CY,          
        cs.MM,          
        cs.HC           
    },
pdfauthor={Aleksandr Poslavsky, Alexander D'yakonov, Yuriy Dorn, Andrey Zimovnov},
pdfkeywords={FRecommender Systems, Industrial Dataset, Short-Video},
}

\begin{document}

\title{VK-LSVD: A Large-Scale Industrial Dataset for Short-Video Recommendation}

\author{
  \href{https://orcid.org/0009-0009-5271-6696}{\includegraphics[scale=0.06]{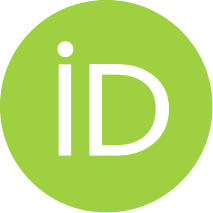}\hspace{1mm} Aleksandr Poslavsky \Letter} \\
  VK AI\\
  VK  \\
  Moscow, Russia\\
  \texttt{dr.slink@vk.com} \\
  \And
  \href{https://orcid.org/0000-0001-7934-6538}{\includegraphics[scale=0.06]{orcid.pdf}\hspace{1mm} Alexander D'yakonov} \\
  VK AI\\
  VK \\
  Moscow, Russia\\
  \texttt{djakonov@mail.ru} \\
  \AND
  \href{https://orcid.org/0000-0003-0533-3018}{\includegraphics[scale=0.06]{orcid.pdf}\hspace{1mm} Yuriy Dorn}\\
  AI Center \& IAI MSU \\
  Lomonosov Moscow State University \\
  Moscow, Russia\\
  \texttt{dornyv@my.msu.ru} \\
  \And
  \href{https://orcid.org/0000-0001-6763-5797}{\includegraphics[scale=0.06]{orcid.pdf}\hspace{1mm} Andrey Zimovnov}\\
  VK AI\\
  VK \\
  Moscow, Russia\\
  \texttt{a.zimovnov@vkteam.ru} \\
}

\maketitle

\begin{abstract}
Short-video recommendation presents unique challenges, such as modeling rapid user interest shifts from implicit feedback, but progress is constrained by a lack of large-scale open datasets that reflect real-world platform dynamics. To bridge this gap, we introduce the \textbf{VK Large Short-Video Dataset (VK-LSVD)}, the largest publicly available industrial dataset of its kind. VK-LSVD offers an unprecedented scale of over \textbf{40 billion interactions} from \textbf{10 million users} and almost \textbf{20 million videos} over \textbf{six months}, alongside rich features including content embeddings, diverse feedback signals, and contextual metadata. Our analysis supports the dataset's quality and diversity. The dataset's immediate impact is confirmed by its central role in the live VK RecSys Challenge 2025.
VK-LSVD provides a vital, open dataset to use in building realistic benchmarks to accelerate research in sequential recommendation, cold-start scenarios, and next-generation recommender systems.\footnote{Accepted to The ACM Web Conference 2026 (WWW '26). 
This is a pre-print version; the final version will appear 
in the conference proceedings.}
\end{abstract}



\date{February 2026}

\section{Introduction}

The rapid expansion of short-video platforms has fundamentally transformed digital content consumption, characterized by frequent, session-based user interactions. Designing effective recommender systems for this domain presents unique challenges, including a reliance on \emph{implicit feedback} (e.g., watch time) as primary signals of user preference and the need to model complex, rapidly evolving user behaviors in conjunction with multi-modal content. Progress is further limited by the scarcity of large-scale, open datasets that faithfully reflect the dynamics and complexity of real-world platforms. Existing public datasets are typically constrained by limited scale, temporal coverage, or feedback diversity, impeding the transfer of academic advances to industrial settings.

To address this gap, we introduce the \textbf{VK Large Short-Video Dataset (VK-LSVD)} \cite{vk_lsvd_2025}, the largest publicly available industrial dataset for short-video recommendation. VK-LSVD is distinguished by its comprehensive feature set, encompassing \textbf{content-based video embeddings}, user socio-demographics, and a wide range of \textbf{implicit} (e.g., watch time), \textbf{explicit} (e.g., likes, dislikes), \textbf{viral} (e.g., shares), and \textbf{deeper engagement} (e.g., comment opens) feedback signals, thus enabling holistic analysis of user behavior. VK-LSVD provides \textbf{key contextual metadata} for each interaction, including consumption context (e.g., feed or search), user platform (e.g.,  Android, Web), and client agent.
Its longitudinal structure and strict \textbf{global temporal split} facilitate robust research on the evolution of user preferences and item popularity. Additionally, VK-LSVD allows researchers to generate custom subsets based on user activity or item popularity, tailored to specific computational or research requirements.


Through data analysis, we validate the dataset’s quality and diversity. Unlike audio content, short videos require active user attention, making every interaction -- whether a full watch or a skip -- a valuable signal for user modeling. VK-LSVD enables the study of these behaviors at an unprecedented scale.

In summary, our contributions are: (1) the introduction of VK-LSVD, (2) a statistical analysis and technical validation of the dataset, (3) a demonstration of its real-world impact as the core dataset for a major recommendation competition.


\section{Related Work}

Public datasets are crucial for recommender systems research, and classics like \textbf{Netflix Prize}~\cite{bennett2007netflix} and \textbf{MovieLens}~\cite{harper2015movielens} are widely used.
However, they offer interactions with limited side information, lacking the sequential nature, high interaction frequency, and multi-modality of modern short-video platforms.

\begin{table*}[t]
\centering 
\caption{Comparison of Modern Short-Video Recommendation Datasets}
\label{tab:dataset_comparison}
\begin{tabular}{lccccccccc}
\toprule
\textbf{Dataset} & \textbf{Platform, Year} & \textbf{Users} & \textbf{Items} & \textbf{Interactions} & \textbf{Temporal Scope}  \\ 
\midrule
VK-LSVD (Ours)                    & VK, 2025 & 10\,M & 20\,M & 40\,B & 6 months  \\ 
Tsinghua U \cite{Shang2025Large} & Kuaishou, 2025 & 10\,K & 153.6\,K & 1\,M & 6 months \\ 
RecFlow \cite{liu2024recflow}    & Kuaishou, 2024 & 42\,K & 82\,M + 9\,M & 38\,M + 1.9\,B & 37 days \\ 
NineRec \cite{zhang2024ninerec}  & Bili, KU, QB, TN, DY, 2024 & 2\,M & 144\,K & 24.5\,M & 10 months \\ 
MovieLens 32M \cite{harper2015movielens} &  MovieLens, 2024 & 200.9\,K & 87.5\,K & 32\,M + 2\,M & 28 years \\ 
MicroLens \cite{ni2023content}   & Micro-video platform, 2023 & 34.5\,M & 1.14\,M & 1\,B & 1 year \\ 
KuaiSAR \cite{sun2023kuaisar}    &  Kuaishou, 2023 & 25.9\,K & 6.9\,M & 19.7\,M & 19 days \\ 
REASONER \cite{Chen2023REASONER} & Custom, 2023 & 3\,K & 4.7\,K & 58.5\,K & - \\ 
KuaiRand \cite{gao2022kuairand}  & Kuaishou, 2022 & 27\,K & 32\,M & 322.3\,M & 2+2 weeks \\ 
KuaiRec \cite{gao2022kuairec}    &  Kuaishou, 2022 & 7.2\,K & 10.7\,K & 12.5\,M & 2 months \\ 
Tenrec \cite{yuan2022tenrec}     & 2 anonymized platforms, 2022 & 5\,M & 3.75\,M & 142.3\,M & 3 months \\ 
\bottomrule
\end{tabular}%
\end{table*}

\subsection{Emerging Short-Video Datasets}

Several Kuaishou-derived datasets address specific needs: \textbf{Kuai\-Rand}~\cite{gao2022kuairand} enables unbiased evaluation; \textbf{KuaiRec}~\cite{gao2022kuairec} offers a near-complete matrix (99.6\% density) with rich side information; \textbf{Kuai\-SAR}~\cite{sun2023kuaisar} integrates search and recommendation behaviors. However, all have limited user scale (up to tens of thousands, see Tab.~\ref{tab:dataset_comparison}) and short observation periods (weeks to months), restricting long-term analysis.

The large-scale \textbf{MicroLens}~\cite{ni2023content} offers raw multi-modal content for end-to-end training but lacks feedback diversity (only comments), global timeline, and uses web-scraped data. The dataset from a \textbf{Tsinghua University} team~\cite{Shang2025Large} integrates user behavior, attributes, and raw video content but at small scale (\textasciitilde 1M interactions).

\subsection{Large-Scale and Multi-Purpose Datasets}

Beyond short-video specifics, 
\textbf{Tenrec}~\cite{yuan2022tenrec} supports cross-domain learning with diverse feedback and true negatives, but lacks raw content. \textbf{NineRec}~\cite{zhang2024ninerec} enables transfer learning via multimodal content across domains, yet has moderate scale, lacks watch time signals, and contains potential biases.

Recent datasets target specialized aspects like interpretation or pipeline consistency.
\textbf{REASONER}~\cite{Chen2023REASONER} provides labeled explanations but is small-scale. \textbf{Rec\-Flow}~\cite{liu2024recflow} supports full-pipeline research with unexposed items, yet has limited users ($\sim$42K), short duration (37 days), and no multi-modal content.

The emergence of large-scale datasets in different domains over the past year highlights the need for more valuable data.
The large-scale \textbf{Yambda-5B}~\cite{ploshkin2025yambda} with 4.79 billion interactions focuses on music recommendation with audio embeddings. The largest \textbf{T-ECD}~\cite{t_tech_ecd_2025} dataset (135 billion cross-domain interactions) offers synthetic e-commerce data for cross-domain learning, but its artificial nature limits real-world applicability, and it does not encompass the short-video format.

Table~\ref{tab:dataset_comparison} reveals a concentration of existing datasets on the Kuaishou platform. To address this lack of platform diversity, we present a novel large-scale dataset from a new ecosystem.

\section{Dataset Description}

This section provides a comprehensive technical description of VK-LSVD's structure, schema, and key statistics to facilitate its use in recommender systems research.
 
\subsection{Data Collection and Ethics}

VK-LSVD is constructed from anonymized interaction logs of the short-video service within VK (the largest social media ecosystems in Russia) -- a high-concurrency environment where users typically engage with dozens of clips per session.

 \paragraph{Anonymization.} All identifiers ( \texttt{user\_id}, \texttt{item\_id}, \texttt{au\-thor\_id}) and categorical context features (\texttt{place}, \texttt{platform}, \texttt{agent}, \texttt{geo}) - are irreversibly anonymized into stable integer IDs. No mapping to the original entities is provided, and all personally identifiable information (PII) has been removed.
\paragraph{Ethical Considerations.} The collection and anonymization processes were designed in compliance with the platform's privacy policy and relevant data protection regulations. \textbf{No raw video, audio, or text content is included}. The provided embeddings are generated via proprietary models, ensuring that the original raw content cannot be reconstructed.

\subsection{Dataset Structure and Key Statistics }

VK-LSVD comprises four core components, with complete data and no missing values across all tables:

\paragraph{Interaction Records (\texttt{interactions/}):} A directory of week\-ly Parquet files (e.g., \texttt{week\_XX.parquet}) containing over 40B user-item exposure events, ordered chronologically, both across files and within each file. Each row represents a unique short-video exposure to a user, encompassing contextual metadata, user feedback signals (watch time, likes, etc.).   The dataset employs a \textbf{Global Temporal Split} (GTS): the \texttt{train} (first 25 weeks, each stored as an individual file), \texttt{validation} (1 week), and \texttt{test} (1 week) subsets represent consecutive time periods.
Each user-item pair in the dataset is unique, recording the parameters of the first interaction; however, the watch time is cumulative across all views, meaning watch time can exceed video duration due to rewatching, despite both being capped (Fig.~\ref{abc}).

\paragraph{User Metadata (\texttt{users\_metadata.parquet}):} Static demographic and behavioral features (age, gender and most frequent geographic location -- 80 unique values) for 10M users, with activity-based sampling support.
\paragraph{Item Metadata (\texttt{items\_metadata.parquet}):} Information for 20M videos including author ID and duration, with popularity-based sampling support.
\paragraph{Embeddings (\texttt{item\_embeddings.npz}):} 
This file provides 64-di\-mensional content-based embeddings for each item. They were initially generated using a local model and subsequently compressed via truncated SVD. Components are ordered by importance to allow for flexible dimensionality trade-offs.

\begin{table}
\centering
\caption{Schema of VK-LSVD}
\label{tab:schema}
\begin{tabular}{p{4.3cm}p{1.9cm}p{5.6cm}}
\hline
\textbf{Field} & \textbf{Type} & \textbf{Description} \\
\hline
\textbf{XX.parquet} & & \\
\hline
user\_id, item\_id & uint32 & Anonymous IDs \\
place, platform, agent & uint8 & Context (feed/search, OS, client) \\
timespent & uint8 & Watch time (capped at 255s) \\
like, dislike, share, bookmark & bool & Feedbacks \\
click\_on\_author & bool & Author profile clicked \\
open\_comments & bool & Comments opened \\
\hline
\textbf{Users Metadata} & &  \textit{(user\_id as above)} \\
\hline
age & uint8 & Age (18-70) \\
gender, geo & uint8 & Gender and location \\
train\_interactions\_rank & uint32 & Popularity rank for sampling \\
\hline
\multicolumn{3}{l}{\textbf{Items Metadata} \ \ \ \ \ \textit{(train\_interactions\_rank, item and user ids as above)}} \\
\hline
duration & uint8 & Duration (seconds) \\
\hline
\textbf{Embeddings} & & \textit{(item\_id as above)} \\
\hline
embedding & float16[64] & Content embedding vector \\
\hline
\end{tabular}
\end{table}

\begin{table} 
\centering
\caption{Core Statistics of VK-LSVD}
\label{tab:core_stats}
\begin{tabular}{lr}
\hline
\textbf{Statistic} & \textbf{Value} \\
\hline
Data Collection Period & 6 months \\
Number of Users & 10,000,000 \\
Number of Items & 19,627,601 \\
Number of Interactions & 40,774,024,903 \\
Total Watch Time & 858,160,100,084 s \\
Dataset Density & 0.0208\% \\

\hline
\textbf{Feedback Type} &  \\
\hline
Number of Likes & 1,171,423,458 \\
Number of Dislikes & 11,860,138 \\
Number of Shares & 262,734,328 \\
Number of Bookmarks & 40,124,463 \\
Clicks on Author & 84,632,666 \\
Comment Opens & 481,251,593 \\

\hline
\end{tabular}
\end{table}

\subsection{Data Availability and Usage}

The VK-LSVD dataset is publicly available on the \textbf{Hugging Face Hub} \cite{vk_lsvd_2025} under the \textbf{Apache License 2.0}, permitting free use for academic and commercial research.
To improve accessibility, we provide \textbf{pre-configured subsets} (e.g., \texttt{ur0.01} for 1\% of random users, \texttt{ip0.01} for 1\% of popular items) and utility scripts to create custom samples that match specific computational resources and  research questions.
For example, we evaluate simple recommendation methods on the \texttt{ur0.01\_ir0.01}-subset (Tab.~\ref{tab:recsys_transposed_clean}): \textbf{Random baseline} (high coverage but no predictive power), \textbf{Global Popularity model} (recommends the most frequently interacted-with items to all users), the \textbf{Conversion-based model} (optimized specifically for predicting click-through rates) and \textbf{iALS} (Implicit Alternating Least Squares --  using watch time $> 10$~s as positive signal, \cite{rendle2022revisiting}).

\begin{table}[h!]
\centering
\caption{Benchmark of Simple Recommendation Algorithms}
\normalsize
\label{tab:recsys_transposed_clean}
\begin{tabular}{lcccc}
\hline
\textbf{Metrics} & \textbf{Random} & \textbf{Popular} & \textbf{Conversion} & \textbf{iALS} \\
\hline
\multicolumn{5}{c}{Global Temporal Split} \\
\hline
Coverage & 0.96449 & 0.00010 & 0.00010 & 0.00501 \\
ROC AUC & 0.50003 & 0.57383 & 0.60341 & 0.58126 \\
NDCG@20 & 0.00006 & 0.00244 & 0.00000 & 0.02623 \\
\hline
\multicolumn{5}{c}{Random Split} \\
\hline
Coverage & 0.97551 & 0.00010 & 0.00010 & 0.00511 \\
ROC AUC & 0.50216 & 0.58999 & 0.68107 & 0.61649 \\
NDCG@20 & 0.00003 & 0.02183 & 0.00000 & 0.06554 \\
\hline
\end{tabular}
\end{table}

The ongoing \textbf{VK RecSys Challenge 2025}, based on this dataset, \cite{vk_challenge_2025} demonstrates VK-LSVD's practical impact. With nearly 800 participants, it tackles a non-standard task: ranking users for new items rather than items for users. Participants are required to predict the top 100 most relevant users for each new, cold-start video, with real-world constraints (max 100 recommendations per user). Evaluation uses NDCG@100 per clip. Results, expected in January 2026, will advance cold-start methods and establish a robust benchmark for sequential and content-based models.


\subsection{Data Analysis}

\begin{figure*}[t]
    \centering
    \includegraphics[width=0.47\textwidth]{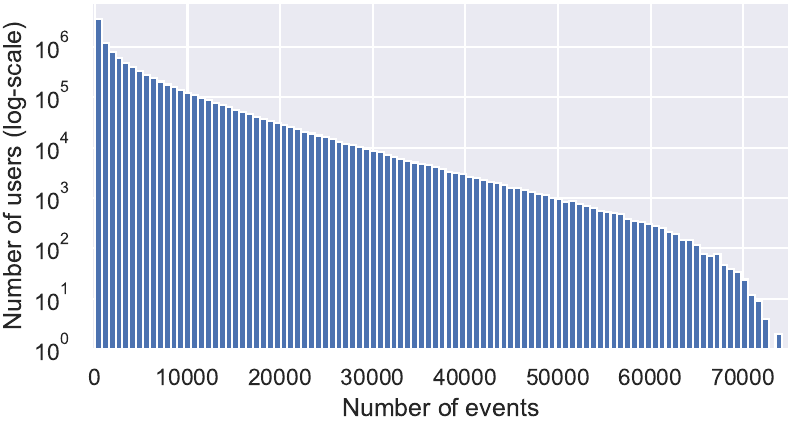}
    \hfill
    \includegraphics[width=0.47\textwidth]{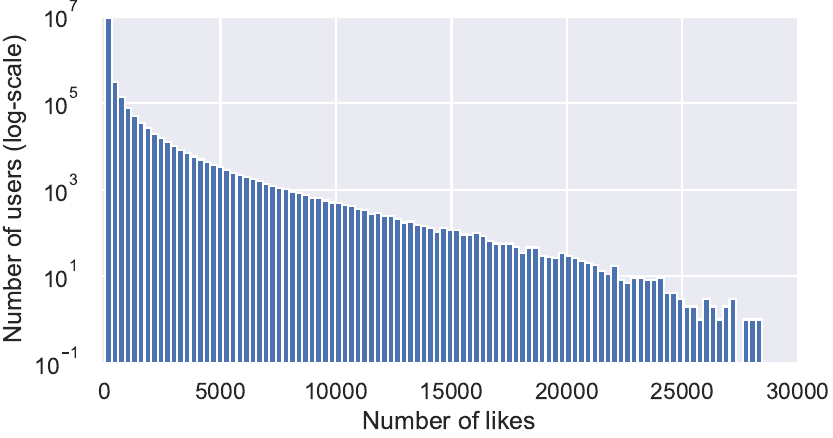}

     \vspace{0.2cm}  

    \includegraphics[width=0.47\textwidth]{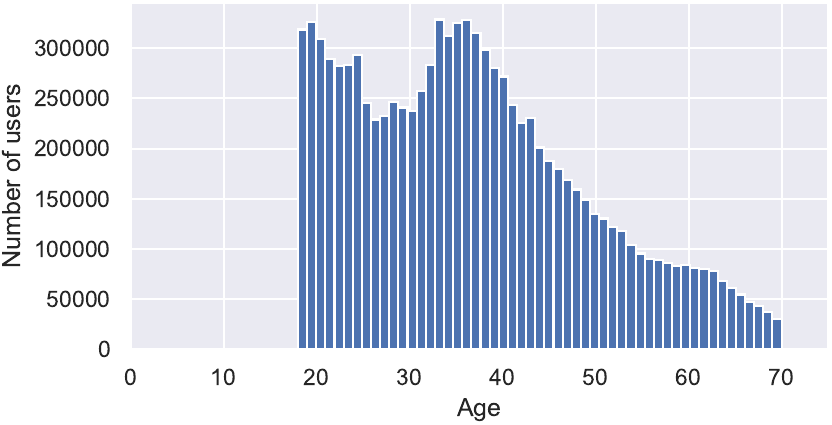}
    \hfill
    \includegraphics[width=0.47\textwidth]{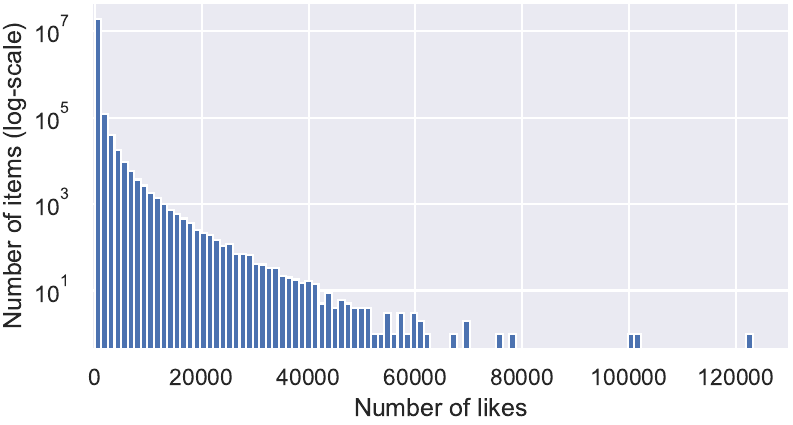}

     \vspace{0.2cm}  

    \includegraphics[width=0.47\textwidth]{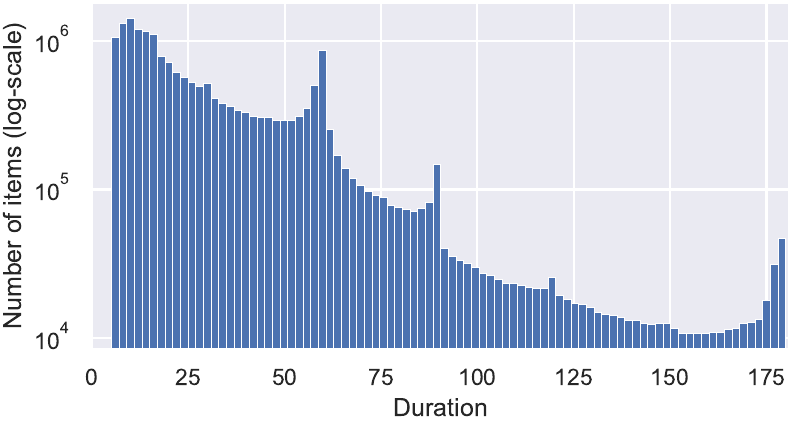}
    \hfill
    \includegraphics[width=0.47\textwidth]{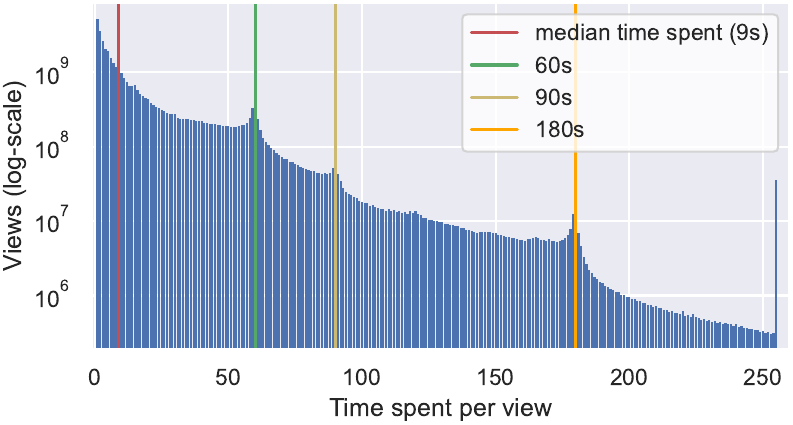}
    \caption{Distributions of Key User and Item Statistics.}
    \label{abc}
\end{figure*}

\textbf{The distribution of user activity} exhibits a highly skewed, heavy-tailed pattern typical of real-world online platforms (Fig.~\ref{abc}): most users generate few viewing events and likes, while a small fraction of ``power users'' accounts for most engagement. Item popularity shows a similar heavy-tailed pattern: a small fraction of items receives the majority of user interactions.

\begin{figure}[t]
\centering
\includegraphics[width=0.69\columnwidth]{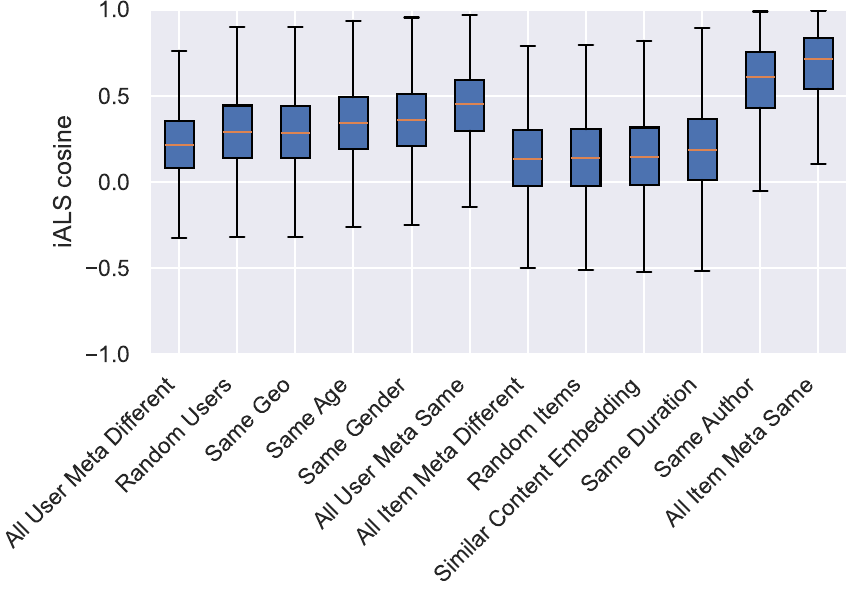} 
\caption{Cosine Similarity Analysis of ALS Latent Factors Across User and Content Attributes}
\label{fig:ALScosines}
\end{figure}

To assess VK-LSVD's utility, we conducted \textbf{a pairwise similarity analysis} using iALS representations \cite{rendle2022revisiting}, trained with a positive target defined as the presence of any positive interaction (like, comment open) and the absence of negative feedback (dislike). User pairs were grouped by demographics (age, gender, location), and item pairs by content features (author, duration, content clusters derived via k-means clustering on the provided 64-D embeddings).
For each group, 10,000 pairs were sampled, and their cosine similarity was computed; random and dissimilar pairs served as baselines.
Results (Fig.~\ref{fig:ALScosines}) show demographic factors (gender, age) strongly influence user similarity, while the author drives item similarity (0.58). The high similarity for items sharing all metadata (0.66) confirms the dataset captures rich content relationships. VK-LSVD enables models to discover meaningful user and content patterns from implicit behavior alone, validating its utility for real-world recommender systems.

\section{Conclusion}

We believe that VK-LSVD will serve as a valuable community resource and accelerate innovation in recommender systems. VK-LSVD is ideally \textbf{suited for research in}:
sequential and session-based recommendation,
next-item prediction,
context-aware and hybrid recommendation,
and the exploration of cold-start and long-tail scenarios.

The expected results of VK RecSys Challenge 2025 will provide an independent and robust evaluation of new approaches developed on VK-LSVD. Technical validation confirms the dataset's value, interaction patterns follow realistic power-law distributions. Furthermore, the provided subsets of the original dataset make it feasible to conduct initial experiments and prototyping on personal workstations, significantly lowering the computational barrier.

While offering demographic attributes, we acknowledge that these features may introduce algorithmic bias. We encourage researchers to use them not only to enhance model performance but also to investigate fairness-aware recommender systems and to mitigate potential disparate impacts across different user groups.\footnote{We thank VK platform users for their anonymized interaction data, and reaffirm our commitment to privacy protection. We acknowledge our colleagues, with special thanks to Dmitry Kondrashkin, Head of VK AI, for his guidance. This work was supported by The Ministry of Economic Development of the Russian Federation in accordance with the subsidy agreement (agreement identifier 000000C313925P4H0002; grant No 139-15-2025-012).}

\bibliographystyle{unsrtnat}  

\bibliography{vkbib}


\end{document}